\documentclass[11pt]{article}

\usepackage{natbib}
\usepackage{amsmath,amsfonts,amssymb,amsthm,amsxtra}
\usepackage{graphicx,psfrag,epsf}
\usepackage{authblk, enumerate}
\usepackage{sectsty}                         
  \allsectionsfont{\usefont{OT1}{phv}{bc}{n}\selectfont}  

\newtheorem{lemma}{Lemma}
\newtheorem{assumption}{Assumption}
\newtheorem{theorem}{Theorem}

\def\be{\begin{equation}}
\def\ee{\end{equation}}
\renewcommand{\theequation}{\thesection.\arabic{equation}}

\newcommand{\argmax}{\operatornamewithlimits{argmax}}
\newcommand{\argmin}{\operatornamewithlimits{argmin}}

\newcommand{\cov}{\text{Cov}}
\newcommand{\var}{\text{Var}}
\newcommand{\E}{\text{E}}

\newcommand{\BGamma}{\pmb{\Gamma}}

\newcommand{\Bc}{{\bf c}}

\newcommand{\BLambda}{\pmb{\Lambda}}

\newcommand{\Btheta}{\boldsymbol{\theta}}

\newcommand{\BA}{{\bf A}}
\newcommand{\Ba}{{\bf a}}
\newcommand{\Bb}{{\bf b}}

\newcommand{\By}{{\bf y}}

\newcommand{\uni}{\text{Uni}}

\title{Discriminant Analysis of Time Series in the Presence of Within-Group Spectral Variability}
\author{Robert T. Krafty}
\affil{  Department of Statistics, Temple University, Philadelphia, PA,   USA  19122 \\ and \\
Department of Biostatistics, University of Pittsburgh, Pittsburgh, PA, USA 15213 \\ rkrafty@pitt.edu }

\date{}

\begin{document}
\maketitle

{\bf Abstract.}
Many studies record replicated time series epochs from different groups with the goal of using frequency domain properties to discriminate between the groups.  In many applications, there exists variation in cyclical patterns from time series in the same group.  Although a  number of frequency domain methods for the discriminant analysis of time series have been explored, there is a dearth of models and methods that account for within-group spectral variability.
This article proposes a model for groups of time series in which transfer functions are modeled as stochastic variables  that can account for both between-group and within-group differences in spectra that are identified from individual replicates.
An ensuing discriminant analysis of stochastic cepstra under this model is developed to obtain parsimonious measures of relative power that optimally separate groups in the presence of within-group spectral variability. The approach possess favorable properties in classifying new observations and can be consistently estimated through a simple discriminant analysis of a finite number of estimated cepstral coefficients. Benefits in accounting for within-group spectral variability are empirically illustrated in a simulation study and through an analysis of gait variability.


\newpage

\section{Introduction}
Discriminant  analysis of time series is important in a variety of fields including physics, geology, acoustics,  economics, and medicine.  In applications where scientifically meaningful information is contained within  power spectra, spectral domain approaches are desired.  A number of spectral based methods for the discriminant and classification analysis of  time series have been developed.  These methods, a review of which can be found in Chapter 7.7 of \cite{shumway2011}, include \cite{shumway1974}, \cite{dargahi1981}, \cite{shumway1982}, \cite{alagon1989}, \cite{zhang1994}, and \cite{kakizawa1998}.

\begin{figure}[h]
\centering
\includegraphics[scale=.9]{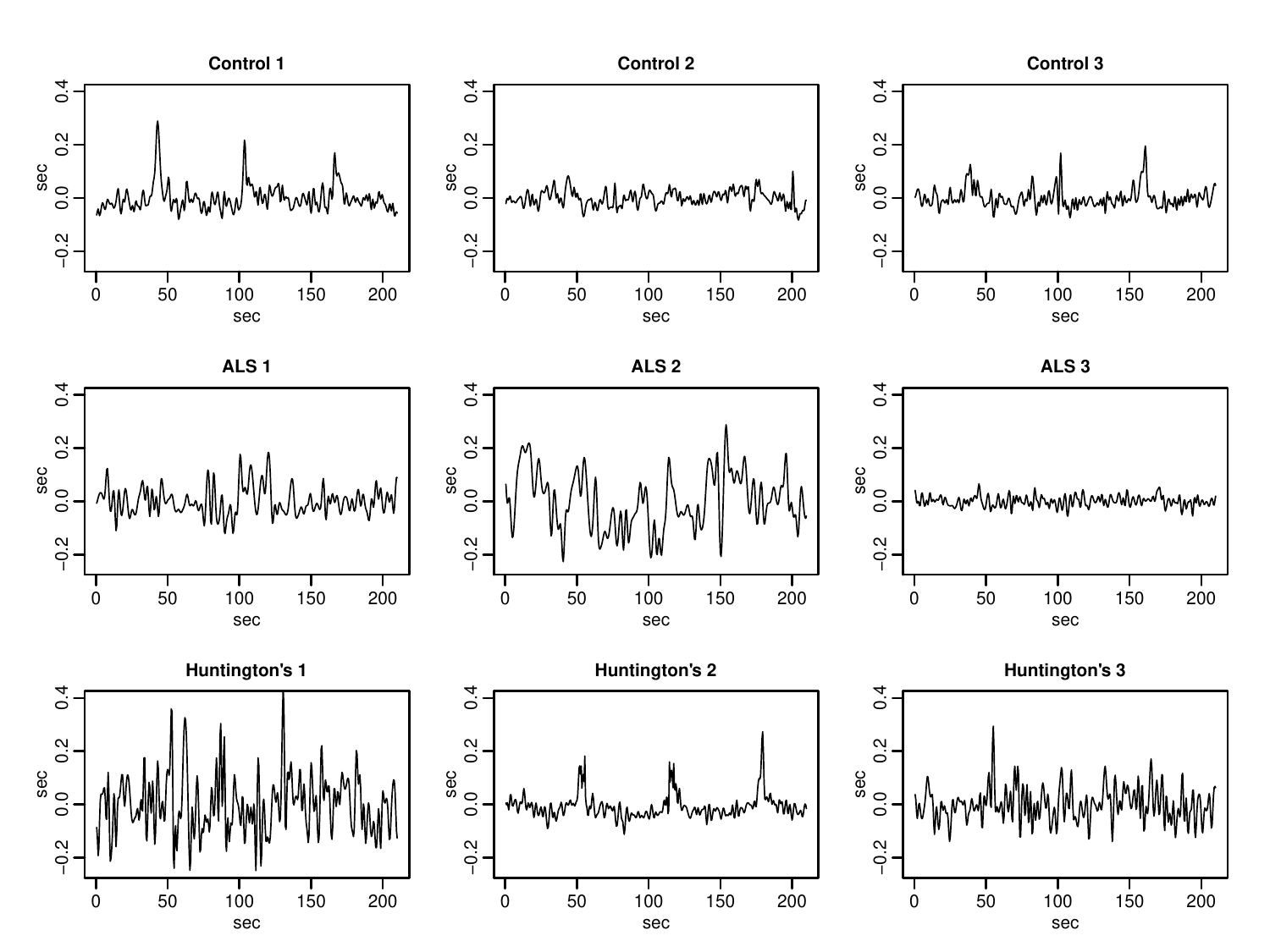}
\caption{\label{fig:rawplot}Detrended stride interval series from 9 participants in the gait analysis study: 3 healthy controls, 3 participants with amyotrophic lateral sclerosis (ALS), and 3 participants with Huntington's disease.}
\end{figure}

\begin{figure}[h]
\centering
\includegraphics{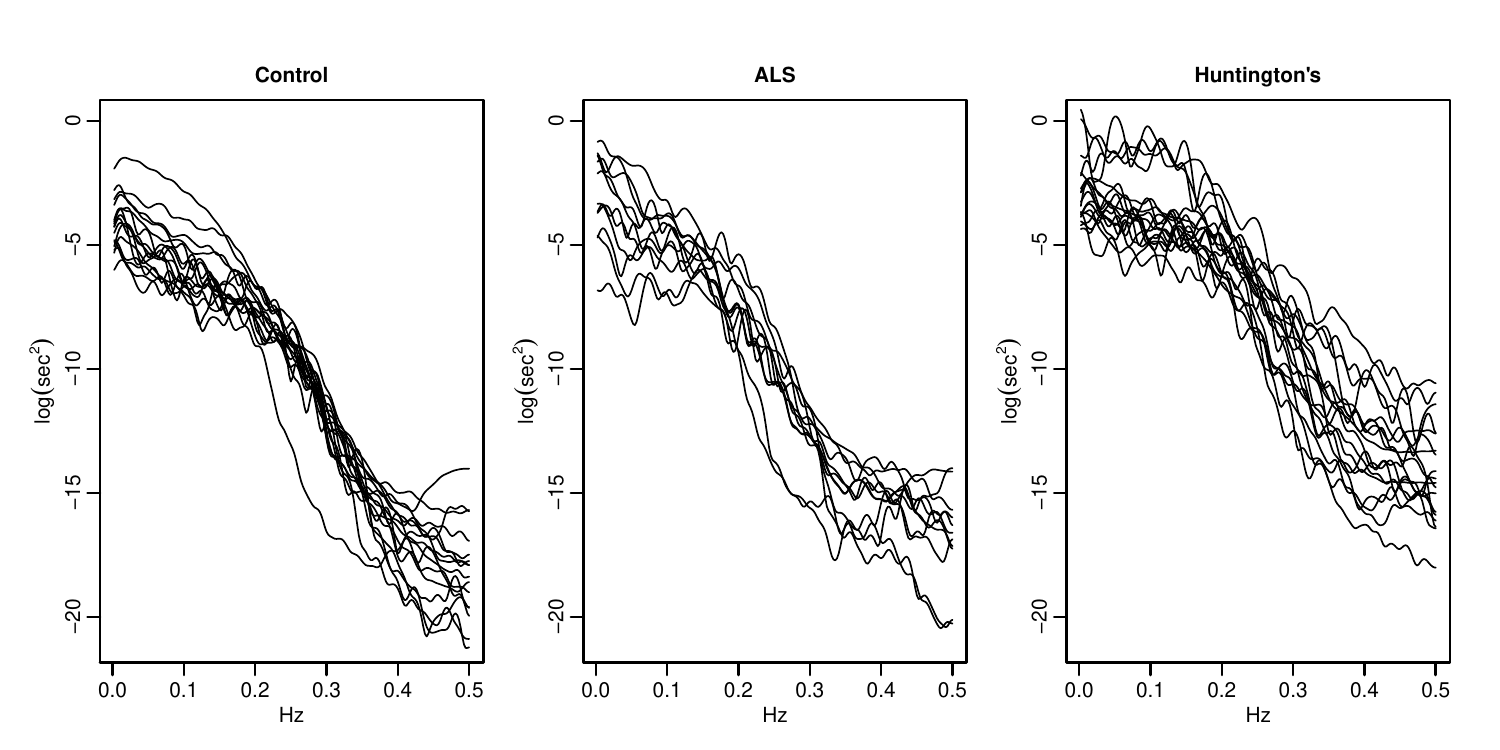}
\caption{\label{fig:rawplotspec}Estimated replicate specific log-spectra for the detrended stride interval series from each participant in the gait analysis study.}
\end{figure}

The aforementioned methods are based on models that assume the existence of group-common power spectra that can be consistently estimated from any single time series replicate from within a group.   In many applications, this assumption does not hold as there exist obvious differences in second-order spectra that can be identified by individual time series within the same group.  As an example, consider stride interval series, or the time taken to complete consecutive gait cycles, that were collected as part of a study to better understand connections between walking patterns and neurological disease \citep{hausdorff2000}.    Figure \ref{fig:rawplot} displays three examples of stride interval series from study participants under each of three neurological conditions and Figure \ref{fig:rawplotspec} displays estimated log-spectra for each stride interval series in the study.   Aside from neurological conditions, walking is influenced by numerous person specific physiological characteristics.  Consequently, there exist differences in cyclical patterns of stride interval series from different subjects with a common neurological condition.

The presence of extra spectral variability in replicated time series and the inability of traditional time series models and methods to account for it was first discussed in the literature by \cite{diggle1997}.  They introduced a parametric log-spectral mixed-effects model for replicated time series, which was later nonparametrically generalized by  \cite{saavedra2000, saavedra2008}, \cite{iannaccone2001}, \cite{freyermuth2010} and   \cite{krafty2011}.  Although these models allow one to conduct inference on a group average spectrum, the question of how within-group spectral variability affects discriminant analysis has yet to be addressed.   To address this question, this article introduces a stochastic transfer function model for groups of time series in the presence of within-group spectral variability that makes explicit the replicate-specific spectra that are estimable from individual series.  The model accounts for the higher-order long-range dependence that is responsible for within-group spectra variability.

The Fisher's discriminant analysis of stochastic cepstra, or inverse Fourier transforms of log-spectra, is explored as a means of discriminating and classifying time series under the stochastic transfer function model.  The procedure possess optimal properties that mirror those of Fisher-type discriminants in other high-dimensional settings.
When extra spectral variability exists, the cepstral Fisher's discriminant analysis more accurately classifies a new time series of unknown group membership compared to existing spectral methods that ignore within-group spectral variability.  Other Fisher's type approaches can be developed under the stochastic transfer function model, such as through integral functions of log-spectra. We use the cepstral formulation as it  provides both a more encompassing theoretical framework and an intuitive approach to estimation that simultaneously overcomes the inconsistency of periodograms and the high-dimensionality of the data.

As discussed in Chapter 11 of \cite{johnson2007}, although intertwined, there is a distinction between discriminant analysis, which seeks parsimonious measures that illuminate group separation, and classification analysis, or the prediction of group membership of new observations.  The aforementioned existing methods for spectral discrimination address the problem of classification, albeit with a higher error than the proposed procedure when within-group variability exists, but are black-box in nature and do little in helping to understand the way in which groups are separated.   The proposed procedure address the discrimination problem by producing discriminants and  weight functions that provide low-dimensional interpretable measures of relative power that illustrate scientific mechanisms that separate groups.

The rest of this article is organized as follows. The stochastic transfer function model for groups of replicated time series with within-group spectral variability is presented in Section \ref{sec:model}.  In Section \ref{sec:lcda}, we discuss the cepstral Fisher's procedure for discriminant and classification analysis. Section \ref{sec:est} introduces an intuitive estimation procedure via a finite number of estimated cepstral coefficients.    Simulation studies are explored in Section \ref{sec:sim} to investigate empirical properties of the proposed procedure and to compare these properties to those of existing methods.   The method is used in Section \ref{sec:gait} to analyze data from the motivating study of gait variability.  Concluding remarks are given in Section \ref{sec:disc}. Proofs are relegated to the appendix.

\section{Model}\label{sec:model}
\subsection{Stochastic Transfer Function Model}\label{sec:model1}
Consider a population of real-valued stationary time series composed of $j=1,\dots,J$  groups, $\Pi_j$ represents the $j$th group, and  $\pi_j$ is the proportion of  time series from the population belonging to $\Pi_j$.   To present a model that accounts for both between and within-group spectral variability, we consider a model with stochastic replicate-specific transfer functions.  Replicate-specific transfer functions $A_{j k}$ are defined as independent second-order random variables identically distributed for each replicate $k$ within group $j$ and whose realizations are complex-valued random functions that are Hermitian, have period 1, and both real and imaginary parts are uniformly continuous.
The group $\Pi_j$ is modeled as the collection of time series  $\left\{X_{j k t} ; t \in \mathbb{Z} \right\}$ where
\begin{equation}\label{eq:stm}
X_{j k t} = \int_{0}^{1} A_{j k}(\lambda)  e^{2 \pi i \lambda t} dZ_{j k}(\lambda),
\end{equation}
$Z_{j k}$ are independent and identically distributed orthogonal processes that are independent of the transfer functions and $\E\left|dZ_{j k}(\lambda)\right|^2 = 1$.  Mixing conditions are assumed on $Z_{j k}$ such that $z_{j k t}= \int_0^1 e^{2 \pi i \lambda t} dZ_{j k}(\lambda)$ is, in some sense, short-range dependent.  Many different mixing conditions can be used.  In this article we use the conditions in Assumption 2.6.1 of \cite{brillinger2001} such that $z_{j k t}$ is strictly stationary, cumulants of all orders exits and are absolutely summable.

Each time series in $\Pi_j$ is an independent and identically distributed stationary process with Cram\'{e}r representation $\int_{0}^{1} e^{2 \pi i \lambda t} dW(\lambda)$, $W(\lambda) = \int_{0}^{\lambda} A_{j k}(\omega)  d Z_{j k}(\omega)$, and spectrum $\E|A_{j k}(\lambda)|^2$.   When $\var\left\{\left|A_{jk}(\lambda)\right|^2\right\} = 0$ for all $\lambda$,  $X_{j k t}$ is short-range dependent.  However, when the variance of replicate-specific spectra is non-trivial,  it is not.  {Conditional} on the replicate-specific random transfer function $A_{j k}$, $X_{j k t}$   is stationary, short-range dependent, and has replicate-specific power spectrum $|A_{j k}(\lambda)|^2$.  The stochastic transfer function model is a reparameterization of the Cram\'{e}r representation that makes explicit the unit-specific spectra that are estimable from individual replicates.  Before comparing the stochastic transfer function model to other models, consider the following example.

\noindent {\bf Example: Conditional MA(1).}
Let $\Pi_j$ be the collection of conditional invertible Gaussian MA(1) processes with nonnegative autocorrelation such that $X_{j k t} = \epsilon_{j k t} + \theta_{j k} \epsilon_{j k t - 1}$ where $\theta_{j k}$ are uniformly distributed over $[0,1]$ and are independent of  $\epsilon_{j k t} \stackrel{iid}{\sim} N(0,\sigma^2)$. These time series have a stochastic transfer function representation of the form (\ref{eq:stm}) where $A_{j k}(\lambda) =  \sigma \left(1 + \theta_{j k} e^{-2 \pi i \lambda}\right)$ and $Z_{j k}$ is complex Brownian motion with $\cov\left\{Z_{j k}(\lambda), Z_{j k}(\omega)\right\} = \min(\lambda, \omega)$ and $Z_{j k}(-\lambda) = \overline{Z_{j k}(\lambda)}$.  { Conditional} on $\theta_{j k}$,  $X_{j k t}$ is a Gaussian MA(1) process with replicate-specific spectrum $\left|A_{j k}(\lambda)\right|^2 = \sigma^2 \left\{1 + \theta_{j k}^2 + 2 \theta_{j k} \cos(2 \pi \lambda) \right\}$.  As a Gaussian MA(1) process, the conditional autocovariance and all conditional higher-order moments vanish for lags greater than 1. Consequently, standard results hold so that the periodogram from a replicate at Fourier frequencies are approximately independent and can be smoothed to obtain a consistent estimate of the replicate-specific spectrum. { Marginally}, $X_{j k t}$ is stationary,  $\cov\left(X_{j k t+h}, X_{j k t}\right)=0$  when $|h|>1$, and it has a marginal power spectrum of $\E \left|A_{j k}(\lambda)\right|^2 = \sigma^2\left\{ 4/3 + \cos(2 \pi \lambda) \right\}$.  However, higher order cumulants do not decay.  For instance, $\cov\left(X^2_{j k t+h}, X^2_{j k t} \right) = 4 \sigma^4/45$ for all $|h| > 1$.  Consequently, periodograms are correlated and cannot be smoothed to obtain a consistent estimate of the marginal spectrum.

The existing models for spectral based discriminant analysis previously cited are based on the assumption that each time series is short-range dependent so that periodograms at different Fourier frequencies are approximately independent, and that periodograms from each realization can be smoothed to obtain a consistent estimate of a  group-common spectrum.  Such assumptions are required in the spectrum analysis of a single time series, as they are essential in obtaining a consistent estimate.  However, they are not appropriate for the analysis of replicated time series when there exists variability in the second-order spectral structures from different realizations.  The stochastic transfer function model generalizes the models considered by these existing methods to account for the marginal higher-order long-range dependence that is responsible for within-group spectral variability.  It should be noted that stochastic transfer functions were previously used in the semiparametric mixed-effects regression model of \cite{krafty2011}, where transfer functions are decomposed into deterministic and stochastic components to allow for the regression analysis of time series when multiple correlated replicates are observed from different subjects.   Although a parameterization of this model using dummy variables can be formulated to define a model of groups of independent time series with extra spectral variability, it is not identifiable.

\subsection{Stochastic Cepstra and Log-Spectra}\label{sec:model2}
Our goal is to find interpretable measures that best separate  groups and focus on those that are measures of relative power, or are linear in log-spectra.    To find such measures, we will utilize cepstral coefficients, or  inverse Fourier transforms of log-spectra. They will  provide both a flexible and rigorous theoretical framework and guide  estimation.   Replicate-specific log-spectra are defined as  $\gamma_{j k}(\lambda) = \log\left|A_{j k}(\lambda)\right|^2 $, the $j$th group mean log-spectrum is defined as  $\alpha_{j}(\lambda) = \E \left\{ \gamma_{j k}(\lambda)\right\}$, and we let $\beta_{j k}(\lambda) = \gamma_{j k}(\lambda) - \alpha_{j}(\lambda)$ be the replicate-specific deviation of the log-spectrum of the $k$th replicate from the $j$th group.   The original formulation of the cepstrum by \cite{bogert1963} considered coefficients in terms of complex trigonometric polynomials. Since we are considering real-valued time series, log-spectra are even functions and, in a manner similar to \cite{bloomfield1973}, we define the cepstrum through a cosine series.   Define the replicate-specific cepstrum as $c_{j k} = \left(c_{j k 0}, c_{j k 1}, \dots \right) \in \mathbb{R}^\mathbb{N}$, where $\mathbb{R}^\mathbb{N}$ is the set of real valued sequences indexed by the natural numbers $\mathbb{N} = \left\{0,1,\dots\right\}$, such that
\begin{equation*}
c_{j k 0} = \int_0^1 \gamma_{j k}(\lambda) d\lambda, \, \, c_{j k \ell} = \int_0^1 \gamma_{j k}(\lambda) \sqrt{2} \cos(2 \pi \lambda \ell) d\lambda,  \, \,  \, \ell=1,2,\dots.
\end{equation*}
Group-average and  replicate-specific deviation cepstra $a_j, b_{jk}   \in \mathbb{R}^{\mathbb{N}}$  are similarly defined as the cosine series of $\alpha_j$ and $\beta_{j k}$, respectively.

\noindent {\bf Example: Conditional MA(1), continued.}  For the collection of conditional MA(1) processes considered in Section \ref{sec:model1},  replicate specific log-spectra are
$$\gamma_{j k}(\lambda) = \log\left[ \sigma^2 \left\{1 + \theta_{j k}^2 + 2 \theta_{j k} \cos(2 \pi \lambda) \right\} \right] = \log(\sigma^2) + 2 \sum_{\ell=1}^{\infty} \frac{(-1)^{\ell + 1} \theta_{j k}^{\ell}}{\ell} \cos(2 \pi \ell \lambda),$$
so that $c_{j k 0} = \log(\sigma^2)$, $c_{j k \ell} = (-1)^{\ell+1} \sqrt{2} \theta^{\ell}_{j k}/ \ell$, $\ell \ge 1$.  Recalling that $\theta_{j k}$ is uniformly distributed over the unit interval,  the group-average cepstra can be found to be $a_{j k 0} = \log\left( \sigma^2 \right)$, $a_{j k \ell} = (-1)^{\ell+ 1} \sqrt{2}/\left\{\ell \left(\ell + 1\right) \right\}$, $\ell \ge 1$.   The zero-order cepstral coefficient reflects the conditional innovation variance.  In our example,  replicates have a common conditional innovation variance, so that $\var\left(c_{j k 0}\right)=0$.  Within-group variability in smooth replicate-specific spectra is reflected in the variability of positive-order cepstral coefficients,  with
\[\cov\left(c_{j k \ell}, c_{j k m} \right) = \frac{2 (-1)^{\ell + m + 2}}{\left(\ell + m + 1\right) \left(\ell + 1\right) \left(m+1\right)}, \, \, \ell, m \ge 1.\]

\section{Discriminant and Classification Analysis}\label{sec:lcda}
\subsection{Cepstral Discriminant Analysis}
The cepstral Fisher's discriminant analysis seeks successive uncorrelated one-dimensional linear functions of replicate-specific cepstra that provide maximum separation between group-means relative to within-group variability.  Before defining the procedure, we first define some notation.  Let $\overline{a}_{\ell} =  \sum_{j=1}^J \pi_{j} a_{j \ell}$ be the overall mean cepstrum and define the between-group kernel as $\Lambda(\ell, m) = \sum_{j=1}^J \pi_j \left(a_{j \ell} - \overline{a}_{\ell} \right) \left(a_{j m} - \overline{a}_{m} \right)$.  Separation between group-means of linear functions of cepstra  $\sum_{\ell = 0}^{\infty} y_{0 \ell} c_{j k \ell}$ with weights $y_{0} \in \mathbb{R}^{\mathbb{N}}$ is defined as the weighed sum of the squared distances of each group-mean to the overall mean, or $\left| \left| y_0 \right| \right|_{\Lambda}^2 = \sum_{\ell,m=0}^{\infty} y_{0 \ell} \Lambda(\ell, m) y_{0 m}$.  Defining the within-group kernel $\Gamma(\ell,m) = E\left(b_{j k \ell} b_{j k m} \right)$, the covariance between two linear combinations of cepstra with weights $y_0, y_1 \in \mathbb{R}^{\mathbb{N}}$ can be written as $\langle y_0, y_1\rangle_{\Gamma} = \sum_{\ell,m =0}^{\infty} y_{0 \ell} \Gamma(\ell,m)  y_{1 m}$ and the variance of a linear combination with weights $y_0$ is given by $\left| \left| y_0 \right| \right|_{\Gamma}^2 = \langle y_0, y_0\rangle_{\Gamma}$.

Discriminants are defined sequentially.  First discriminants $d_{j k 1}$ are linear functions of cepstra defined by weights $y_1 \in \mathbb{R}^{\mathbb{N}}$ such that group means are maximally separated relative to within-group variability such that
\[
d_{j k 1} = \sum_{\ell = 0}^{\infty} y_{1 \ell} c_{j k \ell}, \, \, \, \, \, y_1 = \argmax_{\left|\left|y\right|\right|_{\Gamma} = 1 } \left.  \left\| y \right| \right|_{\Lambda}.
\]
Higher order discriminants maximize group-mean separation among linear combinations orthogonal to lower-order discriminants such that $q$th-order discriminants $d_{j k q}$ with weights $y_q \in \mathbb{R}^{\mathbb{N}}$ are defined as
\[
d_{j k q} = \sum_{\ell = 0}^{\infty} y_{q \ell} c_{j k \ell},  \, \, \, \, \, y_q = \argmax_{\left|\left|y\right|\right|_{\Gamma} = 1, \, \,  \langle y, y_m \rangle_{\Gamma} = 0, m<q } \left.  \left\| y \right| \right|_{\Lambda}.
\]
The number of non-trivial discriminants $Q$ is less than or equal to the ranks of $\Gamma$ and $\Lambda$, which is less than or equal to $J-1$.  The discriminants $d_{j k q}$ provide low-dimensional measures that best separate the $J$ groups.  These parsimonious measures can be used to visualize the high-dimensional data, such as Figure 5 in the analysis of gait variability in Section \ref{sec:gait}, and provide an intuitive a powerful classification procedure, which is discussed in Section 3.3. The weights $y_q$ provide information as to how the discriminants can be interpreted as measures of relative power.

The cepstral Fisher's discriminant analysis is formulated above when the covariance functions of $b_{j k}$ are the same for each group.  If there is heterogeneity, all statements made in this article aside from those concerning optimal classification rates still apply using the pooled within-group covariance $\sum_{j=1}^J \pi_j \E(b_{j k \ell} b_{j k m})$ in lieu of  $\Gamma(\ell,m)$.  A more detailed discussion of this issue is presented in Section \ref{sec:disc}. A consequence of Proposition 2.1 from \cite{shin2008} is that cepstral discriminants are well defined when cepstral group means are not dissimilar to replicate-specific deviations in the sense that $a_j$ is contained in the reproducing kernel Hilbert space with reproducing kernel $\Gamma$ for $j=1,\dots,J$.  We assume that this regularity condition is satisfied so that discriminants  always exist.

\subsection{Log-Spectral Weight Functions and Discriminants}\label{sec:logspec}
Measures of relative power could have alternatively been defined using integral functions of log-spectra. However, due to the unbounded nature of the inverse of a non-singular continuous covariance operator, optimal discriminants will not necessarily exist unless additional strong and nonassessable assumptions are made.  This issue is discussed by  \cite{shin2008} and by \cite{delaigle2012}.  When  $\sum_{\ell=0}^{\infty} |y_{q \ell}| < \infty$, ${\xi}_q(\lambda) = {y}_{q 0} + \sum_{\ell=1}^{\infty} {y}_{q \ell} \sqrt{2} \cos(2 \pi \lambda \ell)$ exists, $d_{j k q} = \int_0^1 \xi_q(\lambda) \gamma_{j k}(\lambda) d \lambda$, and we refer to $\xi_q$ as a log-spectral weight function.  The cepstral based formulation is broader and encompass the integral log-spectral formulation in the sense that, if a discriminant exists in the integral log-spectral formulation, it is equivalent to $d_{j k q}$ and has weight function $\xi_q$.

\subsection{Classification Analysis}\label{sec:class}
Consider a new time series of unknown group membership  $\left\{ X_{* t}; t \in \mathbb{Z} \right\}$ with cepstrum $c_{*}$ and $q$th cepstral discriminant $d_{* q}$.  The property that $d_{* q}$, $q=1,\dots,Q$, are uncorrelated with unit variance suggests classifying the new time series into $\Pi_j$ when the $j$th group-mean  discriminants  $\mu_{j q} = \sum_{\ell=0}^{\infty} y_{q \ell} a_{j \ell}$ are closer to $\left(d_{*1},\dots,d_{*Q} \right)^T$  than the other $J-1$ group means. Formally, if we let $\Pi(c_*) \in \left\{1,\dots,J\right\}$  index  the population to which the new time series is classified, the classification rule is
\begin{equation}\label{eq:class}
\Pi(c_*) = \argmin_{j=1,\dots,J} \left\{ \sum_{q=1}^Q \left(d_{* q} - \mu_{j q} \right)^2 -2\log(\pi_j)\right\}.
\end{equation}

This classification rule is the optimal centroid classifier of \cite{delaigle2012} for replicate-specific log-spectra and  properties concerning its classification rate are established in their Theorems 1 and 2  when $\Gamma$ is non-singular and $J=2$.  If the processes generating  $\gamma_{j k}$ are Gaussian, then (\ref{eq:class}) is  optimal in that it has smallest classification error among all spectrum based classification rules, and this error is bounded from zero.  If $\gamma_{j k}$ is not Gaussian, then the classification rule with smallest error will depend on the distribution of $\gamma_{j k}$, and this error is bounded from zero.
Although a non-linear classifier will exist with smaller classification error,  one must know the distribution of $\gamma_{j k}$ to find this rule and, while it presents some theoretical improvements compared to the Fisher's cepstral procedure for the classification problem, it will not provide parsimonious measures to address the discrimination problem.

When within-group spectral variability is not present, asymptotically perfect classification can be achieved \citep{zhang1994, kakizawa1998}.  However, when within-group spectral variability is present, asymptotically perfect classification is not possible and methods that can achieve asymptotically perfect classification in the absence of within-group spectral variability have a bounded non-zero error rate.  As a time series from one group could possess a replicate-specific spectrum that more closely resembles replicate-specific spectra from another group by chance alone, this is rather intuitive and illustrates the increased difficulty of classification in the presence of within-group spectral variability.  It should be noted that \cite{delaigle2012} describe a situation where asymptotically perfect classification is possible for functional data.  However, the assumption that $a_j$ is contained in the reproducing kernel Hilbert space with reproducing kernel $\Gamma$, which is known to be necessary in avoiding degenerate time series models \citep{parzen1961, parzen1962}, makes asymptotically perfect classification not possible under the stochastic transfer function model.

\section{Estimation}\label{sec:est}

\subsection{Cepstral Coefficients}
Consider  estimation  from $n=\sum_{j=1}^J n_j$ independent time series epochs of length $N$, $\left\{X_{j k 1}, \dots, X_{j k N} \right\}$, $j=1,\dots,J$, $k=1,\dots,n_j$.  When replicate-specific log-spectra are smooth,  most information is contained in lower-order cepstral coefficients.   This is illustrated in the conditional MA(1) example from Sections \ref{sec:model1} and \ref{sec:model2}, where $c_{j k \ell} = O_p\left(\ell^{-3/2}\right)$. Consequently, discriminants and weight functions can be estimated through a classical Fisher's discriminant analysis of a finite number of estimated cepstral coefficients.

We consider the class of replicate-specific cepstral estimators of the form
\begin{equation*}
\hat{c}_{j k 0} = N^{-1} \sum_{m=0}^{N-1} \hat{\gamma}_{j k m}, \, \, \hat{c}_{j k \ell} = N^{-1} \sum_{m=0}^{N-1} \hat{\gamma}_{j k m} \sqrt{2} \cos\left(2 \pi \lambda_{m} \ell \right), \,\,\, \ell=1,\dots,\lfloor N/2 \rfloor,
\end{equation*}
where $\hat{\gamma}_{j k m}$ is an estimator of $\gamma_{j k}(\lambda_{m})$, $\lambda_{m} = m/N$, $m \in \mathbb{Z}$.
There is an extensive literature on spectrum estimation.  Since, for fixed $\ell$, the variance of $\hat{c}_{j k \ell} - c_{j k \ell}$ decays as $N$ increases, not only are standard consistent estimators appropriate, such as smoothed periodograms  and multitaper estimates, but also inconsistent estimators, such as periodograms.  Although the consistency results for discriminants and weight functions established in Section \ref{sec:con} hold for any standard estimator, good finite sample performance is contingent on log-spectral estimators having small bias, small variance and, since the within-group covariance $\Gamma$ must also be estimated, errors that are approximately uncorrelated across frequency.  We advocate the use of the multitaper estimators developed by \cite{thomson1982} that are shown in \cite{percival1993} to
have small bias, as opposed to periodogram based methods, small variance, as opposed to inconsistent estimators, and high resolution, as opposed to methods that smooth across frequency.

Let $\left\{h_{r t}, t=1,\dots,N \right\}$ be $r=1,\dots,R$ non-negative orthonormal data tapers such that $\sum_{t=1}^N h_{r t} h_{s t} = 1\left\{r=s\right\}$, $r,s = 1, \dots, R$, where $1\left\{\cdot\right\}$ is the indicator function.    This article will consider the sine tapers
\begin{equation*}
h_{r t} = \left(\frac{2}{N+1}\right)^{1/2} \sin\left( \pi t \frac{r}{N+1} \right).
\end{equation*}
The $r$th direct spectral estimator is defined as the tapered periodogram under the $r$th data taper, $I_{j k r m} = \left|N^{-1/2} \sum_{t=1}^N h_{r t} X_{j k t} e^{-2 \pi i \lambda_{m} t} \right|^2$, and the multitaper log-spectral estimator is defined as
\begin{equation}\label{eq:mtap}
\hat{\gamma}_{j k m} = \log\left( R^{-1} \sum_{r=1}^R I_{j k r m} \right).
\end{equation}

\subsection{Discriminants, Weight Functions, and Classification}\label{sec:est:2}
We estimate discriminants and weight functions through a classical Fisher's discriminant analysis of $\hat{\Bc}^L_{j k} = \left(\hat{c}_{j k 0}, \dots, \hat{c}_{j k L-1}\right)^T$ for some $L$ smaller than $N$ and $n$.  The data driven selection of $L$ is discussed in Section \ref{sec:K}.   To define this estimator, let $\hat{\pi}_j = n_j/n$, $\hat{\Ba}^L_{j} = n_j^{-1} \sum_{k=1}^{n_j} \hat{\Bc}^L_{j k}$, $\hat{\overline{\Ba}}^L = \sum_{j=1}^J \hat{\pi}_j \hat{\Ba}^L_j$, and $\hat{\BLambda}_L = \sum_{j=1}^J \hat{\pi}_j \left(\hat{\Ba}^L_j - \hat{\overline{\Ba}}^L\right)\left(\hat{\Ba}^L_j - \hat{\overline{\Ba}}^L\right)^T$.  Further,  let $\hat{\Bb}^L_{j k} = \hat{\Bc}^L_{j k} - \hat{\Ba}^L_{j}$ and   $\hat{\BGamma}_{L} = \sum_{j=1}^J \hat{\pi}_j \hat{\BGamma}_{L j}$ where $\hat{\BGamma}_{L j} = (n_j - 1)^{-1}  \sum_{k=1}^{n_j} \hat{\Bb}^L_{j k}  \left(\hat{\Bb}^L_{j k}\right)^T$.
An estimated $q$th cepstral weight function $\hat{\By}_q^L$ is defined as a $q$th ordered eigenvector of $\hat{\BGamma}_L^{-1} \hat{\BLambda}_L$ and an estimated discriminant is defined as $\hat{d}^L_{j k q} = \left(\hat{\By}_q^L\right)^T \hat{\Bc}_{j k}^L$. Note that the estimated log-spectral weight function $\hat{\xi}^L_q(\lambda) = \hat{y}^L_{q 0} + \sum_{\ell=1}^{L-1} \hat{y}^L_{q \ell} \sqrt{2} \cos(2 \pi \lambda \ell)$  exists and is interpretable even if $\xi_q$ itself only exists in a limiting sense.  The classification rule for a new time series with  estimated $L$--dimensional cepstrum $\hat{\Bc}^L_*$ can be estimated as $\hat{\Pi}(\hat{\Bc}^L_*) = \text{argmin}_{j} \left[ \sum_{q=1}^Q \left\{ \left(\hat{\Bc}_*^L-\hat{\Ba}_j^L \right)^T \hat{\By}^L_q \right\}^2 - 2 \log\left(\hat{\pi}_j\right) \right]$.


It should be noted that $\hat{c}_{j k \ell} \approx c_{j k \ell}$ for all $\ell =0,\dots,L-1$ when a reasonable spectral estimator is chosen and when $L$ is small relative to $N$.  Consequently, in practice, the procedure is similar to a classical Fisher's discriminant analysis on $\left(c_{j k 0}, \dots, c_{j k L-1}\right)^T$, and statistical properties of estimators follow from classical results for estimated Fisher's discriminant analysis \citep[Chapter 6]{anderson2003}.

\subsection{Selecting $L$}\label{sec:K}
A leave-out-one cross-validation procedure akin to the cross-validation commonly used for other regularized Fisher's discrimination procedures can be used for the data driven selection of $L$. 
 Let $\hat{\Pi}^L_{[j k]}$ be the index of the classification rule for the $k$th time series from $\Pi_j$ using $L$ cepstral coefficients when  weight functions are estimated using the $n-1$ time series that exclude this series.  The cross-validation rule selects
$L =  \text{argmin}_{\ell} \left[ \sum_{j=1}^J \sum_{k=1}^{n_j} 1\left\{ \hat{\Pi}^{\ell}_{[j k]} \ne j \right\} \right]$.

\subsection{Consistency}\label{sec:con}
Consistency of estimated discriminants, weight functions, and classification rules can be established under some assumptions as $n_j$, $N$, and  $L$ increase.   First, it is assumed that $\hat{\pi}_j$ is $\sqrt{n}$-consistent, which holds under simple random sampling.

\begin{assumption}
\normalfont
$\hat{\pi}_j = \pi_j + O_p\left(n^{-1/2}\right).$
\end{assumption}

\noindent Appropriate estimators of $\pi_j$ based on the sampling scheme can be used if this does not hold.

We will assume the following two regularity conditions so that cepstral coefficients decay at a sufficient rate and that eighth moments of $X_{j k t}$ exist.

\begin{assumption}
\normalfont
$\sum_{\ell=1}^{\infty} \ell^2 \left|a_{j \ell} \right|^2 < \infty$ and $\Pr\left( \sum_{\ell=1}^{\infty} \ell^2 \left|b_{j k \ell} \right|^2 < \infty \right) = 1$, $j=1,\dots,J$.
\end{assumption}

\begin{assumption}
\normalfont
$\sup_{\lambda \in \mathbb{R}} E \left\{| dZ_{j k}(\lambda)|^8\right\} < \infty$ and $\sup_{\ell \in \mathbb{N}} \E \left|b_{j k \ell}\right|^4 < \infty$,  $j=1,\dots,J$.
\end{assumption}

\noindent Assumption 2 implies that mean and replicate-specific deviations of log-spectra are absolutely continuous with square integrable first derivatives and, consequently, error incurred by using only a finite number of cepstral coefficients is asymptotically negligible.   The moment conditions in Assumption 3 imply that the eighth moments of $X_{j k t}$ are bounded.  Our estimators depend on $\hat{\BGamma}_L$ and $\hat{\BLambda}_L$, which are fourth-order functions of $X_{j k t}$, and these conditions assure their variances exist.   Additionally, it is assumed that a log-spectral estimator is chosen that has asymptotically optimal bias (up to a constant), resolution, and bounded variance in the sense that the following assumption holds.

\begin{assumption}
\normalfont
For $j=1,\dots,J$, $\sup_{m=0,\dots,\lfloor N/2 \rfloor}\E \left\{\hat{\gamma}_{j k m} - \gamma_{j k}(\lambda_m) \right\} = \nu_1 + O(N^{-1/2}),$ for some $\nu_1 \in \mathbb{R}$,
$\sup_{m\ne p=0,\dots,\lfloor N/2 \rfloor}\cov\left\{\hat{\gamma}_{j k m} - \gamma_{j k}(\lambda_m), \hat{\gamma}_{j k p} - \gamma_{j k}(\lambda_p) \right\} = O(N^{-1}),$  and \\
$\sup_{m=0,\dots,\lfloor N/2 \rfloor}\var\left\{\hat{\gamma}_{j k m} - \gamma_{j k}(\lambda_m)\right\} = \nu_2 + O(N^{-1})$ for some $\nu_2 > 0 $.
\end{assumption}

\noindent Log-spectral estimators need only be asymptotically unbiased up to a constant since the discriminant analysis is invariant to the addition of a constant.  Assumption 4 holds for multitaper estimates, including tapered periodograms when $R=1$.  Whereas consistency results for multitaper estimates from a single realization require the number of tapers to grow with respect to the number of time points, we consider fixed $R$, since the variance reduction achieved by increasing the number of tapers is inherently achieved by projecting onto a finite cosine basis.

\begin{lemma}\label{lemma0}
Assumption 4 holds for $\hat{\gamma}_{j k m}$ defined by $(\ref{eq:mtap})$ under Assumptions 2 and 3.
\end{lemma}

\noindent The proof of Lemma \ref{lemma0} mirrors the proof of Theorem 1 from \cite{krafty2011}.  It should be noted that, although this theorem is formulated for untapered periodograms, the properties underlying the Bartlett's expansion used in the proof that are formulated in \cite{janas1995} hold for tapered data.

The following theorem establishes consistency under these assumptions when the growth of $L$ is restricted by the growth of $n$ and $N$. The main obstacle in performing Fisher's discriminant analysis when the dimension is large compared to the number of observations stems from the divergence of the eigenvalues of the within-group covariance and its inverse. The smoothness of log-spectra implies that the largest eigenvalue of the truncated within-group covariance matrix $\E\left\{ \Bb^L_{j k} \left(\Bb^L_{j k}\right)^T \right\}$ is bounded from above but that its smallest, which we will refer to as $\sigma_L$, approaches zero as $L \rightarrow \infty$.
To assure that the inverse of  $\E\left\{ \Bb^L_{j k} \left(\Bb^L_{j k}\right)^T \right\}$ can be consistently estimated by its sample version, we must assume that the number of replicates grows quickly compared to the number of coefficients by assuming that $L n^{-1/2} \rightarrow 0$ and $\sigma^{-2}_L n^{-1/2} \rightarrow 0$.  Replicated-specific cepstral coefficients are not observed but estimated from replicates of length $N$.  To ensure that the error incurred by this estimation is asymptotically negligible,  we must also limit the growth of the number of coefficients by the length of the time series such that $\sigma^{-2}_{L} N^{-1/2} \rightarrow 0$.

\begin{theorem}
Under Assumptions 1-4, for every $q$th weight function $y_q$, there exist a series of $q$th eigenvectors $\hat{\By}^L_q$ of
$\hat{\BGamma}_L^{-1} \hat{\BLambda}_L$ such that, if $\sigma^{-2}_L n^{-1/2} \to 0$, $\sigma^{-2}_L N^{-1/2} \to 0$ and $L n^{-1/2} \to 0$  as $n,N,L \rightarrow \infty$, $\left| \left| \left(\hat{\By}_q^L, 0, \dots \right) - y_q \right| \right|_{\Gamma} \overset{p}{\to} 0$.
\end{theorem}

A direct consequence of the $\Gamma$--norm consistency of weight functions is that $\hat{d}^L_{j k q} \overset{p}{\to} d_{j k q}$ and  $\hat{\Pi}(\hat{\Bc}^L_*) \overset{p}{\to} \Pi(c_*)$.

\section{Simulation}\label{sec:sim}
\begin{figure}[t]
\centering
\includegraphics{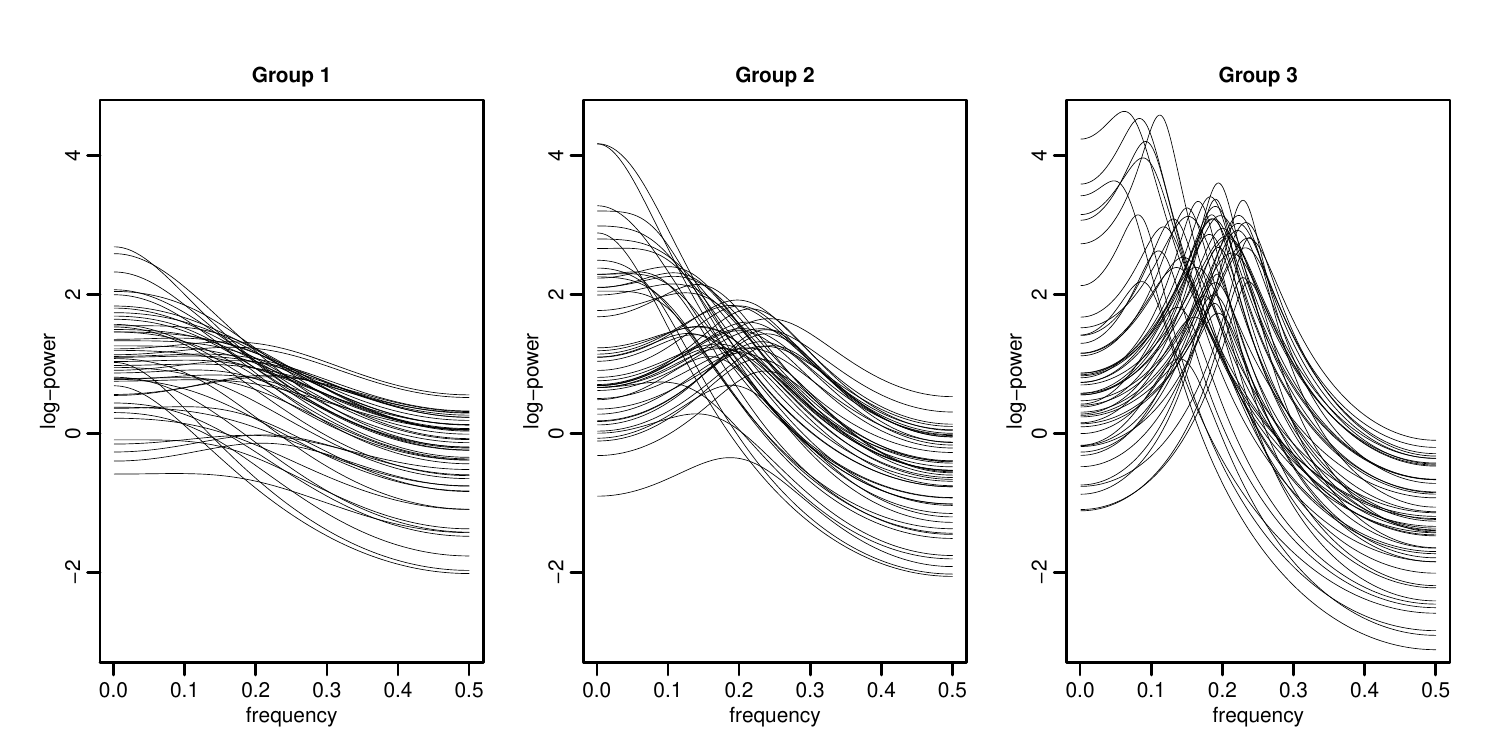}
\caption{\label{fig:simlogspec} Replicate-specific log-spectra from simulated conditional AR(2) processes with $n_j=50$ and  $\sigma^2_{j k} \in \left[0.3, 3\right]$.}
\end{figure}

We conducted a simulation study to investigate the empirical properties of classification procedures when within-group spectral variability exists.
In these simulations, time series from $J=3$ groups are simulated as conditional autoregressive processes $X_{j k t} = \phi_{j k 1} X_{j k t-1} + \phi_{j k 2} X_{j k t-2} + \epsilon_{j k t}$ where $\epsilon_{j k t}$ are independent Gaussian white noise with conditional variance $\sigma^2_{j k}$.   Autoregressive parameters  are drawn as independent uniform random variables where $\phi_{1 k 1} \sim \uni\left(0.05,0.7\right)$, $\phi_{1 k 2} \sim \uni\left(-0.12,-0.06\right)$,  $\phi_{2 k 1} \sim \uni\left(0.01,1.2\right)$, $\phi_{2 k 2} \sim \uni\left(-0.36, -0.25\right)$,  $\phi_{3 k 1} \sim \uni\left(0.12,1.5\right)$, and $\phi_{3 k 2} \sim \uni\left(-0.75, -0.56\right)$.  Three ranges for conditional innovation variances are explored where $\sigma^2_{j k}$ are drawn as independent uniform random variables over $[0.1, 10]$, $[0.3, 3]$, and $[0.9, 1.1]$. One thousand random samples are drawn for each of the 27 combinations of the three ranges of innovation variances, three numbers of time series per group in training data $n_j=15, 50, 100$, and three time series lengths $N=250, 500, 1000$.  Figure \ref{fig:simlogspec} displays simulated replicate-specific log-spectra when $n_j=50$ and $\sigma^2_{j k} \in \left[0.3, 3\right]$.  A test data set of 50 time series per group is drawn for each random sample to evaluate out-of-sample classification rates.

The proposed cepstral Fisher's discriminant analysis was implemented using three log-spectral estimators:  the multitaper estimator with $R=7$, a direct estimator, and a smoothed estimator.   Direct estimators were taken to be first tapered log-periodograms.  Smoothed  estimates were obtained by taking the logarithm after smoothing $I_{j k 1 m}$ across frequency via the modified Daniell smoother with span selected through generalized cross-validation \citep{ombao2001b}. Additionally, two popular methods for the discriminant analysis of time series in the absence of within-group spectral variability were implemented using Kullback-Leibler and Chernoff information measures  \citep{kakizawa1998, shumway2011}.   Replicate-specific spectra were individually estimated by smoothing periodograms with a modified Daniell smoother and span selected through generalized cross-validation.  A test time series is classified into $\Pi_j$ when the information measure distance between its smoothed periodogram and the average of the smoothed periodograms from the $\Pi_j$ training data is smaller than its distance to the average of the smoothed periodograms from the training data from the other two groups.  The tuning parameter for the Chernoff measure was selected using an appropriately modified version of the cross-validation procedure outline in Section \ref{sec:K}.

    \begin{table}\small
    \centering
\renewcommand\arraystretch{.75}
\begin{tabular}{l l l c c c c c}
                &      &       & Cepstral        &  Cepstral       &  Cepstral          &                  &  Kullback-\\
$\sigma^2_{j k}$&$n_j$ & $N$    & Multitaper      & Direct          & Smoothed        & Chernoff         & Leibler  \\
$[0.1, 10]$     &15    & 250    & 91.1 (3.5)    & 87.3 (3.7)    & 89.4 (3.6)    & 80.2 (4.9)     & 79.2 (4.9) \\
                &      & 500    & 94.0 (2.7)    & 91.5 (3.4)    & 93.2 (2.7)    & 81.1 (4.8)     & 80.1 (4.6)\\
                &      & 1000   & 95.6 (2.8)    & 94.1 (3.1)    & 95.4 (2.7)    & 81.4 (4.8)     & 80.7 (4.7) \\
                &50    & 250    & 93.1 (2.2)    & 90.0 (2.6)    & 91.8 (2.5)    & 83.0 (3.5)     & 81.9 (3.4) \\
                &      & 500    & 95.5 (1.8)    & 93.6 (2.2)    & 94.9 (1.9)    & 84.1 (3.8)     & 82.8 (3.7)\\
                &      & 1000   & 97.0 (1.7)    & 95.9 (1.7)    & 96.8 (1.6)    & 84.7 (3.6)     & 83.5 (3.6) \\
                &100   & 250    & 93.8 (2.0)    & 90.6 (2.4)    & 92.3 (2.2)    & 84.2 (3.3)     & 82.8 (3.5) \\
                &      & 500    & 95.7 (1.7)    & 93.9 (2.0)    & 95.2 (1.8)    & 84.8 (3.2)     & 83.5 (3.3)\\
                &      & 1000   & 97.2 (1.5)    & 96.0 (1.6)    & 96.9 (1.5)    & 85.3 (3.2)     & 84.2 (3.5) \\
$[0.3, 3]$      &15    & 250    & 91.1 (3.1)    & 86.8 (4.3)    & 89.3 (3.7)    & 82.7 (3.9)     & 82.4 (4.0) \\
                &      & 500    & 93.9 (3.0)    & 91.6 (3.6)    & 93.4 (3.0)    & 83.7 (3.9)     & 83.2 (4.0)\\
                &      & 1000   & 95.5 (2.7)    & 94.1 (2.9)    & 95.4 (2.7)    & 84.4 (4.1)     & 83.9 (4.1) \\
                &50    & 250    & 93.1 (2.1)    & 89.8 (2.7)    & 91.7 (2.3)    & 84.9 (3.2)     & 84.4 (3.4) \\
                &      & 500    & 95.5 (1.8)    & 93.7 (2.1)    & 94.9 (1.9)    & 85.8 (3.4)     & 85.3 (3.5)\\
                &      & 1000   & 97.0 (1.6)    & 95.7 (1.8)    & 96.8 (1.6)    & 86.6 (3.0)     & 85.9 (3.5) \\
                &100   & 250    & 93.7 (2.0)    & 90.5 (2.5)    & 92.3 (2.2)    & 85.4 (2.9)     & 84.8 (3.0) \\
                &      & 500    & 95.7 (1.7)    & 94.0 (2.0)    & 95.2 (1.8)    & 86.5 (2.9)     & 86.0 (3.1)\\
                &      & 1000   & 97.3 (1.5)    & 96.1 (1.7)    & 97.0 (1.5)    & 87.4 (2.8)     & 86.8 (3.1) \\
$[0.9, 1.1]$    &15    & 250    & 91.2 (3.3)    & 86.9 (4.0)    & 89.6 (3.6)    & 85.9 (3.2)     & 85.7 (3.2) \\
                &      & 500    & 93.9 (3.0)    & 91.5 (3.4)    & 93.4 (3.0)    & 86.6 (3.2)     & 86.3 (3.3)\\
                &      & 1000   & 95.8 (2.6)    & 94.4 (2.7)    & 95.5 (2.6)    & 87.6 (3.1)     & 87.3 (3.2) \\
                &50    & 250    & 93.2 (2.2)    & 89.9 (2.6)    & 91.8 (2.3)    & 86.2 (3.0)     & 86.1 (3.0) \\
                &      & 500    & 95.5 (1.8)    & 93.6 (2.2)    & 95.0 (1.9)    & 87.1 (2.8)     & 86.6 (3.0)\\
                &      & 1000   & 97.1 (1.6)    & 95.7 (1.8)    & 96.8 (1.6)    & 88.1 (2.8)     & 87.6 (3.0) \\
                &100   & 250    & 93.6 (2.0)    & 90.4 (2.3)    & 92.3 (2.2)    & 86.2 (2.9)     & 85.8 (3.0) \\
                &      & 500    & 96.0 (1.7)    & 94.1 (2.0)    & 95.4 (1.8)    & 87.6 (2.6)     & 87.0 (2.9)\\
                &      & 1000   & 97.3 (1.4)    & 96.1 (1.6)    & 97.1 (1.5)    & 88.4 (2.6)     & 88.0 (2.9) \\
\end{tabular}
        \caption{\label{tab}Mean (standard deviation) of the percent of correctly classified replicates.}
\end{table}

Table 1 displays the mean and standard deviation of the classification rates.  In every setting, the cepstral Fisher's discriminant analysis, using any of the three log-spectral estimators, had higher mean classification rates than the two information criterion based methods.   Although the three cepstral Fisher's discriminant procedures display comparatively similar performance, in each setting, the multitaper based method had the best classification rate and the direct method has the poorest.  Changes in the amount of within-group spectral variability by changing the range of conditional innovation variances, which is the same for each group, do not affect the performance of the cepstral Fisher's procedures.  However, the classification rates of the procedures that ignore within-groups spectral variability have reduced classification rates when this variability increases.

\section{Analysis of Gait Variability}\label{sec:gait}
 Patterns of gait variability can provide insight into how neurological conditions  affect the systems that regulate walking \citep{hausdorff2000, hausdorff2005}.  The discriminant analysis  of gait variability from people with different neurological conditions can provide a tool for characterizing pathologies,  aid in the diagnosis of neurodegenerative disease, and aid in the evaluation of treatment efficacy.  In this section, we consider a discriminant analysis of gait variability from three groups of participants in the the study  described by \cite{hausdorff2000}:  healthy controls, participants with amyotrophic lateral sclerosis, a disease characterized by a loss of motoneurons,  and participants with Huntington's disease, a pathology of the basal ganglia.  These data can be obtained through {\tt PhysioNet} \citep{goldberger2000}.

\begin{figure}
\centering
\makebox{\includegraphics[width=5in]{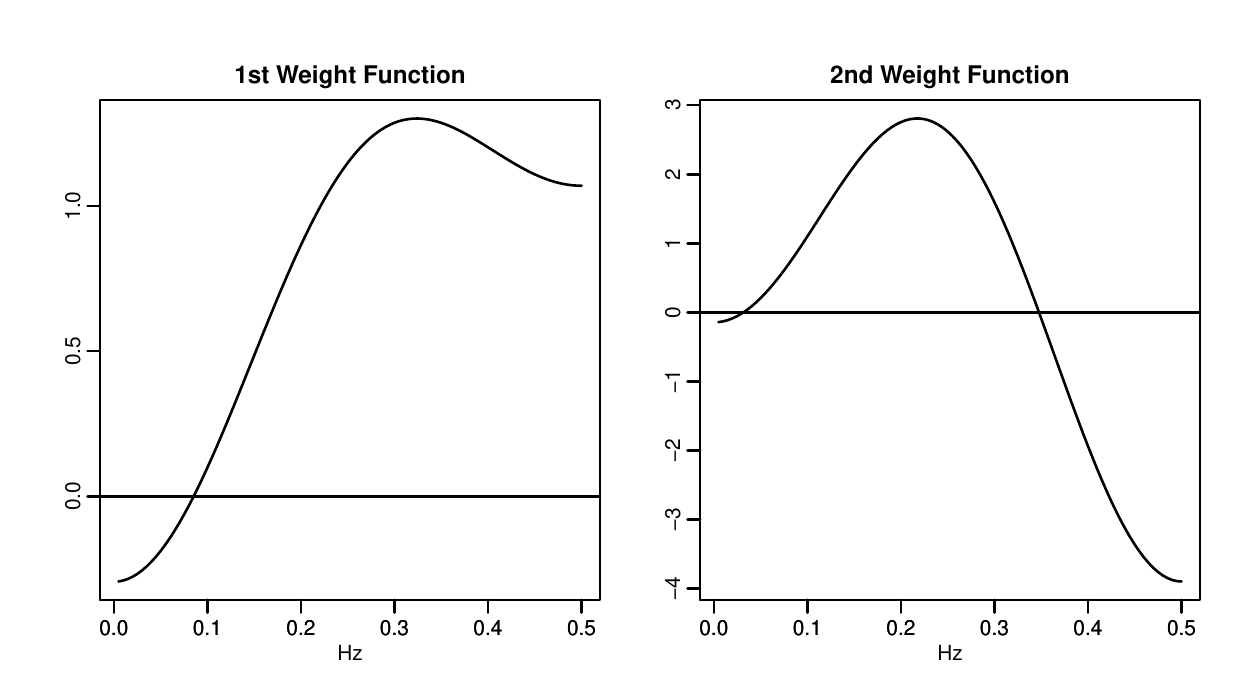}}
\caption{\label{fig:weightFuncs}Estimated log-spectral weight functions from the gait analysis study.}
\end{figure}

In the study, participants were fitted with pressure sensors on the soles of their feet and told to walk at a normal pace.  The information collected was used to compute stride intervals, or the elapsed time for each gait cycle.  The present analysis considers the 3.5 minutes of stride intervals defined by the left foot  after a 20 second start-up period.   After a 3 standard deviation median filter was applied to remove artifacts associated with turning at the end of the hallway \citep{hausdorff2000}, cubic smoothing spline interpolants of the stride intervals as  functions of time were sampled at 2 Hz and linear trends were removed.  The resulting data are detrended stride interval series of length $N=420$ from $n=45$
participants:  16 healthy controls, 11  participants with amyotrophic lateral sclerosis, and 18  participants with Huntington's disease.

\begin{figure}
\centering
\makebox{\includegraphics[width=4.75in]{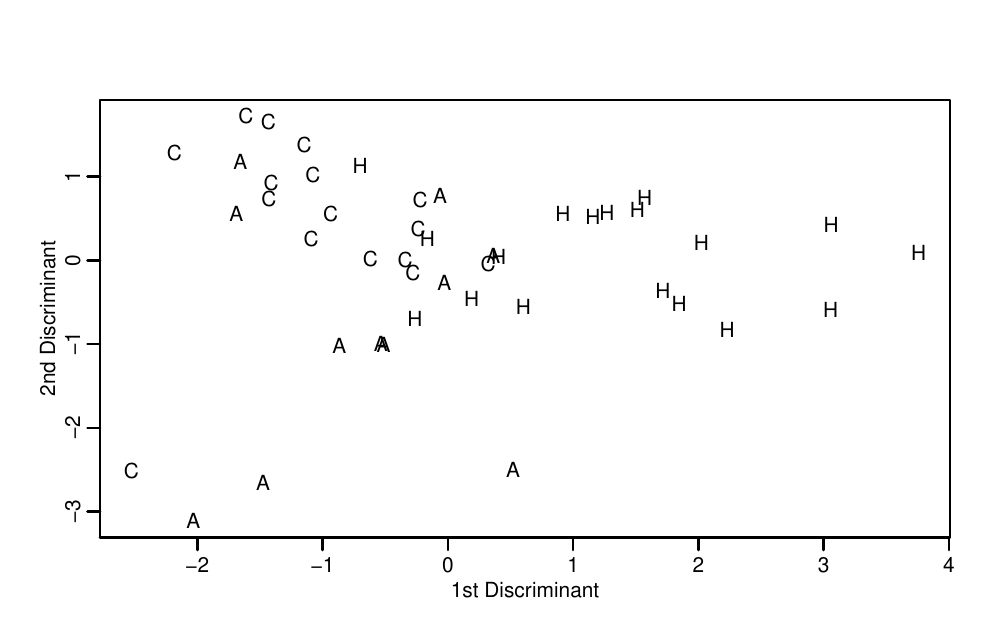}}
\caption{\label{fig:discPlot}Estimated discriminants from the gait analysis study: ``C'' denotes healthy controls, ``A'' denotes participants with amyotrophic lateral sclerosis, and ``H'' denotes participants with Huntington's disease.}
\end{figure}

Log-spectra were estimated using $R=7$ multitapers and cross-validation selected $L=4$ coefficients.    Estimated log-spectral weight functions $\hat{\xi}_1$ and $\hat{\xi}_2$ are displayed in Figure \ref{fig:weightFuncs}.
To aid interpretability, we only display power spectra for frequencies between $[0, 0.5]$ Hz \citep{henmi2009}.

The first log-spectral weight function is a contrast in power from  frequencies greater than $0.08$ Hz versus power from  frequencies less than $0.08$ Hz;  large positive values for the first discriminant indicate larger relative power from higher frequencies.  The second weight function is a contrast between power from frequencies between 0.05--0.35 Hz versus power from frequencies greater than 0.35 Hz; large positive values for the second discriminant indicate more power from lower frequencies.  A scatter plot of the estimated discriminants $\hat{d}_{j k 1}, \hat{d}_{j k 2}$ is displayed in Figure \ref{fig:discPlot}. The first discriminant primarily separates participants with Huntington's disease from the other two groups while the second discriminant primarily separates controls from participants with amyotrophic lateral sclerosis.  This finding that  groups are separated by contrasts in power from higher frequencies versus lower frequencies is not unexpected.  Neurodegenerative diseases such as amyotrophic lateral sclerosis and Huntington's disease affect systems that are important in maintaining a steady stride,  thus introducing power at higher frequencies.  \cite{hausdorff2000} found that, on average, positive values of the autocorrelation function decay slower for controls as compared to participants with Huntington's disease, which is indicative of healthy controls possessing more power at lower frequencies, and that the average decay for participants with amyotrophic lateral sclerosis is between that of controls and participants with Huntington's disease.

Leave-out-one cross-validation was used to empirically assess the effectiveness of the classification rule.  Fourteen of the 16 controls were correctly classified,  1 misclassified as having amyotrophic lateral sclerosis, and 1 misclassified as having Huntington's disease.  Fifteen of the 18 participants with Huntington's disease were correctly classified, 2 incorrectly classified as being controls, and 1 misclassified as having amyotrophic lateral sclerosis.  The rule did comparatively worse in classifying participants with amyotrophic lateral sclerosis:  5 were correctly classified, 4 classified as controls, and 2 classified as having Huntington's disease.  The Kullback-Leibler and Chernoff information measure  classifiers of \cite{kakizawa1998} were also implement and their performances evaluated though leave-out-one cross-validation.  The two information measure classifiers had identical performance. They correctly classified 11 of the 16 controls,  12 of the 18 participants with Huntington's disease, and 5 of the 11 participants with amyotrophic lateral sclerosis.  In addition to providing a more accurate classification rule,  cepstral Fisher's discriminant analysis produced interpretable estimated log-spectral weight functions and two-dimensional discriminant plot that can be used for illustration, which is lacking from the information measure classifiers.

\section{Discussion}\label{sec:disc}
Traditional models for the frequency domain discriminant analysis of time series assume each series from a group are independently and identically distributed as models that are used in the spectral analysis of a single time series.  However, such models fail to account for within-group spectral variability that is present in most real world applications.  This article introduced the stochastic transfer function model that can account for this within-group variability, which is a product of long-range higher-order dependence.  A cepstral based Fisher's discriminant analysis is developed under this model.  The procedure provides parsimonious low-dimensional measures that illuminate scientific mechanisms that best separate groups and, when within-group spectral variability is present, provides more accurate classification of new observations as compared to methods that do not account for within-group spectral variability.

The cepstral Fisher's discriminant analysis was presented in this article under the assumption of equal within-group spectral covariance functions.  If these covariance functions differ  such that $\Gamma_j(\ell,m) = E\left(b_{j k \ell} b_{j k m}\right)$ depends on $j$, the discriminants can be defined by using $\Gamma = \sum_{j=1}^J \pi_j \Gamma_j$.  Under this setting, if group membership is viewed as a random variable such that a randomly selected time series has probability $\pi_j$ from being in group $j$, discriminants have the interpretation of being linear functions that maximize the variance of the conditional expected value relative to the expected value of the conditional variance, where conditioning is taken with respect to group membership.  The proposed estimator, which already utilizes the estimate $\hat{\BGamma}_L = \sum_{j=1}^J \hat{\pi}_j \hat{\BGamma}_{L j}$, is still valid in this setting and the consistency results established in Theorem 1 hold. The only statements made in this article that need to be adjusted when log-spectral covariances depends on group are those that are made with respect to classification rates.  The optimal classification rule under normality in this settings will be quadratic and, while providing some gain in classification accuracy, fails to address the discriminant analysis question or provide interpretable parsimonious measures, such as the estimated weight functions and discriminants in Figures \ref{fig:weightFuncs} and \ref{fig:discPlot}.  Additionally, although the theoretically optimal classifier is quadratic, Fisher's classification procedures are robust to heteroscedasticity \citep{oneill1992}  and outperform quadratic procedures in practice for high-dimensional data where parameters must be estimated \citep{cheng2004}.

Only univariate time series were considered in this article while the information criteria based methods of \cite{kakizawa1998} are applicable to vector-valued time series.  Although the stochastic transfer function model for time series in the presence of within-group spectral variability can be extended to the vector-valued  setting in a straight forward manner, due to the non-convexity of the matrix exponential discussed in Section 6 of \cite{krafty2013},  multivariate analogues of cepstral coefficients have yet to be developed.  Consequently, the extension of the proposed discriminant analysis procedure to the vector-valued setting is not straight forward.  A cepstral discriminant analysis could be conducted on the collection of component-specific cepstral coefficients, however such an analysis ignores coherence between different components.  The extension of the proposed procedure to the vector-valued setting will be the focus of future work.

\section*{Acknowledgements}
This work was supported by  grant R01GM113243 from the National Institute of General Medical Sciences, U.S.A.  I thank two anonymous revivers whose insightful comments greatly improved the article.

\setcounter{equation}{0}
\renewcommand{\theequation}{A.\arabic{equation}}

\section*{Appendix:  Proofs}
\subsection*{Proof of Theorem 1}\label{sec:proof}
To prove Theorem 1, we will make use of two lemmas.  Lemma \ref{lemma:1}, whose proof is provided in the subsequent subsection, formalizes the approximation of the true infinite dimensional weight functions from the weight functions obtained from truncated true cepstra $\Bc^L_{j k} = \left(c_{j k 0}, \dots, c_{j k L-1} \right)^T$, while Lemma \ref{lemma:2} follows directly from Assumptions 2-4 and the orthonormality of the cosine series.

\begin{lemma}\label{lemma:1}
Under Assumption 2,   for a $q$th cepstral weight function $y_q$, there exists a series of $q$th  weight functions $\By_q^L$ for the Fisher's discriminant analysis of $\Bc^L_{j k}$ such that
$ \left| \left| \left(\By^L_q,0,\dots\right) - y_q \right| \right|_{\Gamma} \to 0$ as $L \rightarrow \infty$
\end{lemma}

\begin{lemma}\label{lemma:2}
Under Assumptions 2-4, as $N \rightarrow \infty$,  $E\left(\hat{c}_{j k 0} - c_{j k 0} - \nu_1 \right)^2 =  O\left(N^{-1}\right)$ and \\ $\sup_{\ell=1,\dots,\lfloor N/2 \rfloor} E \left( \hat{c}_{j k \ell} - c_{j k \ell} \right)^2 = O\left(N^{-1}\right)$.
\end{lemma}

Define $\Ba^L_{j} = \left(a_{j 0}, \dots, a_{j  L-1} \right)^T$, $\Bb^L_{j k} = \left(b_{j k 0}, \dots, b_{j k L-1} \right)^T$, $\BGamma_L = \E\left\{ \Bb^L_{j k} \left( \Bb^L_{j k}\right)^T \right\}$, and $\BLambda_L = \sum_{j = 1}^J \pi_j \left(\Ba^L_{j} - \overline{\Ba}^L \right) \left( \Ba^L_{j} - \overline{\Ba}^L\right)^T$ where $\overline{\Ba}^L = \sum_{j=1}^J \pi_j \Ba^L_j$.
Further, let  $|| \cdot ||$ be the operator norm on $L \times L$ matrices such that, for an $L \times L$ matrix $\BA$, $||\BA|| = \sup_{\By^T \By = 1} \left(\By^T \BA^T \BA \By\right)$.  Note that showing $||\hat{\BGamma}_L^{-1} \hat{\BLambda}_L - \BGamma_L^{-1} \BLambda_L ||\overset{p}{\to} 0$ implies that $\hat{\By}^L_q \overset{p}{\to} \By^L_q$, and since Lemma \ref{lemma:1} established that $\By^L_q$ converges to $y_q$ in $\left| \left| \cdot \right| \right|_{\Gamma}$, completes the proof of Theorem 1.  To ease exposition, we simplify the notation in this proof by suppressing the dependence of parameters and estimates on  $L$.
From Cauchy-Schwarz  and basic matrix algebra, it follows that
\begin{eqnarray}\label{eq:eig}
 \left|\left|\hat{\BGamma}^{-1} \hat{\BLambda} - {\BGamma}^{-1} {\BLambda} \right|\right|
 &\le& \left| \left| {\BGamma}^{-1} \left( \hat{\BGamma} - \BGamma \right) \hat{\BGamma}^{-1}  \hat{\BLambda}  - \BGamma^{-1} \left( \hat{\BLambda} - \BLambda \right) \right| \right|  \nonumber \\
 &\le& \sigma^{-1} \left| \left| \hat{\BGamma} -\BGamma \right| \right| \left| \left|\hat{\BGamma}^{-1} \right| \right| \left| \left| \hat{\BLambda} \right| \right|+ \sigma^{-1} \left| \left| \hat{\BLambda} - \BLambda \right| \right|.
 \end{eqnarray}

To investigate the decay of $|| \hat{\BGamma} - \BGamma || \le \sum_{j=1}^J | | \hat{\BGamma}_j - \BGamma ||$, note that, since the operator norm is dominated by the Frobenius norm, $\E | | \hat{\BGamma}_j - \BGamma ||^2 \le \sum_{\ell,m=0}^{L-1} \E | \hat{\Gamma}_j(\ell,m) - \Gamma(\ell,m) |^2$.    From Lemma \ref{lemma:2} and the definition of $\hat{\BGamma}_j$, letting  $\eta_{\ell} = \sup_{j} \E|b_{j k \ell}|^4$, there exists a $C > 0$ such that $\E \left\{ \hat{\Gamma}_j(\ell,m) - \Gamma(\ell,m) \right\}^2 \le C \sqrt{\eta_\ell \eta_{m}} \left(n^{-1} + N^{-1}\right)$, so  $\E | | \hat{\BGamma}_j - \BGamma ||^2 \le C \left(\sum_{\ell=0}^{L - 1} \sqrt{\eta}_{\ell} \right)^2 \left(n^{-1} + N^{-1} \right)$.  Since $\Pr\left( \sum_{\ell=1}^{\infty} \ell^2 \left|b_{j k \ell} \right|^2 < \infty \right) = 1$ and $\E \left|b_{j k \ell}\right|^4 < \infty$, $\eta_{\ell} = O(\ell^{-4})$ and  $\lim_{L \rightarrow \infty} \left( \sum_{\ell=0}^{L-1} \sqrt{\eta}_{\ell}\right)^2 < \infty$.  It then follows that $|| \hat{\BGamma} - \BGamma || = O_p\left(n^{-1/2}\right) + O_p\left( N^{-1/2} \right)$.

It follows from similar arguments to those used above that $| | \hat{\BLambda} - \BLambda | | = O_p\left(n^{-1/2}\right) + O_p\left( N^{-1/2} \right)$.  Since the approximation of the finite between-group kernel is stable and the infinite dimensional kernel is bounded, $|| \BLambda | | = O_p(1)$ \citep[Proposition 2.2]{chatelin1981}, and consequently, $|| \hat{\BLambda} | | = O_p(1)$.

To investigate the decay of $||\hat{\BGamma}^{-1}||$, let $\hat{\sigma}$ be the smallest eigenvalue of $\hat{\BGamma}$ so that $||\hat{\BGamma}^{-1} || = \hat{\sigma}^{-1}$.  Define the matrix $\tilde{\BGamma} = \BGamma^{-1/2} \hat{\BGamma} \BGamma^{-1/2}$, let $\tilde{\sigma}$ be its smallest eigenvalue, and  note that $\hat{\sigma}^{-1} \le \sigma^{-1} \tilde{\sigma}^{-1}$.  It follows from Lemma \ref{lemma:2} and from \cite{bai1993} that $\tilde{\sigma} = 1 + O_p(L n^{-1}) + O_p(N^{-1/2})$, and consequently when $N \to \infty$ and $L n^{-1/2} \to 0$, $||\hat{\BGamma}^{-1}|| = \sigma^{-1} + O_p(\sigma^{-1}L n^{-1}) + O_p(\sigma^{-1}N^{-1/2})$.

Plugging these results for $|| \hat{\BGamma} - \BGamma ||$, $||\hat{\BLambda} - \BLambda||$, $||\hat{\BLambda}||$ and $||\hat{\BGamma}^{-1} ||$ into Equation \ref{eq:eig}, we conclude that  $||\hat{\BGamma}^{-1} \hat{\BLambda} - {\BGamma}^{-1} {\BLambda} || = O_p(\sigma^{-2} n^{-1/2}) + O_p(\sigma^{-2} N^{-1/2}) + O_p(\sigma^{-2} L n^{-1})$ which, when $\sigma^{-2}n^{-1/2} \to 0$, $\sigma^{-2}N^{-1/2} \to 0$ and $Ln^{-1/2} \to 0$, decays to zero.

\subsection*{Proof of Lemma 2}\label{sec:lemma1}
For an $\mathbb{R}^{\mathbb{N}}$--valued random variable $g
$ with covariance kernel $\Gamma$, let  $L^2(g)$ represent the Hilbert space spanned by linear functions of $g$ with inner product $\langle \sum_{\ell=0}^\infty w_{\ell} g_{\ell}, \sum_{m=0}^\infty v_{m} g_{m} \rangle_{g} = \sum_{\ell,m=0}^{\infty} w_{\ell} v_{m} \Gamma(\ell,m) = \langle w, v \rangle_{\Gamma}$.  Further, let $\langle \cdot, \cdot \rangle _{\cal H}$ be the inner product of the reproducing kernel Hilbert space ${\cal H}(\Gamma)$ that has reproducing kernel $\Gamma$.  As discussed in \cite{aronszajn1950}, there exists a congruence between ${\cal H}(\Gamma)$ and $L^2(g)$ defined through the linear map $\Psi \left\{ \Gamma(\ell,\cdot) \right\} = g_\ell$.  Define the operator $T: {\cal H}(\Gamma) \rightarrow {\cal H}(\Gamma)$ as $\left(T h \right)_{\ell} = \langle \Lambda(\ell, \cdot), h \rangle_{\cal H}$, $h \in {\cal H}(\Gamma)$.  The operator $T$ is self-adjoint, positive and compact so that it possess a spectral decomposition $T = \sum_{q=1}^Q \tau_q \theta_q \otimes \theta_q$ where $\tau_q$ are nonincreasing eigenvalues and $\theta_q$ are orthonormal eigenfunctions in ${\cal H}(\Gamma)$ \citep[Theorem 2.4]{shin2008}. We will also consider $\tilde{\Lambda}_L$ as the minimum ${\cal H}(\Gamma) \otimes {\cal H}(\Gamma)$ norm interpolate to $\Lambda_L$ on $(0,\dots,L-1)\times(0,\dots,L-1)$ and its associated operator $T_L:{\cal H}(\Gamma) \rightarrow {\cal H}(\Gamma)$ defined as $\left(T_L h \right)_{\ell} = \langle \tilde{\Lambda}_L(\ell,\cdot), h \rangle_{\cal H}$.  Let $\tau^L_q$ be the nonincreasing set of eigenvalues of the operator $T_L$ and $\Btheta^L_q$ be a set of associated eigenfunctions.
The following lemma, whose proof mirrors the proof of Lemma 4 in \cite{eubank2008} that deals with approximating the operator of a canonical correlation analysis, relates the spectral decomposition of $T_L$ to the Fisher's discriminant analysis of $\Bc^L_{j k}$.

\begin{lemma}\label{lem:thm1}
For every $q$th eigenfunction $\theta^L_q$ of $T_L$, there exists a $q$th eigenfunction $\By^L_q$ of $\BGamma_L^{-1} \BLambda_L$ such that $|| \Psi(\theta^L_q) - \sum_{\ell=0}^{L-1} y^L_{q \ell} g_\ell ||_{g} = 0$.
\end{lemma}
A consequence of Lemma \ref{lem:thm1} and the $L^2(g)$ equivalence between $\Psi\left(\theta_q\right)$ and $\sum_{\ell=0}^{\infty} y_{q \ell} g_{\ell}$  is that the proof of Lemma 2 can be completed by showing the existence of $q$th eigenfunctions $\theta^L_q$ of $T_L$ such that $\theta^L_q \rightarrow \theta_q$ in ${\cal H}(\Gamma)$.

By Cauchy-Schwarz, $|| \left(T-T_L \right) h ||_{\cal H} \le ||\Lambda - \tilde{\Lambda}_L||_{{\cal H} \otimes {\cal H}} ||h||_{\cal H}$ for all $h \in {\cal H}(\Gamma)$ where $|| \cdot |||_{{\cal H} \otimes {\cal H}} $ is the tensor product norm.  The decay of cepstral coefficients under Assumption 2 implies that  $||\Lambda - \tilde{\Lambda}_L ||_{{\cal H} \otimes {\cal H}} \rightarrow 0$ and consequently, $T_L$ converges to $T$ in operator norm.  This convergence, when combined with the property that $T_L$ is stable, implies that the set of unique eigenvalues of $T_L$ converges to the set of unique eigenvalues of $T$ \citep[Proposition 2.2]{chatelin1981}.   Further, if $\tau^L_q$ is a series of eigenvalues that converge to $\tau_q$,  the multiplicity of $\tau^L_q$ is always greater than or equal to the multiplicity of $\tau_q$ \citep[Lemma 2.1]{chatelin1981}.  Since there exists a $L_0$ such that the rank of $T_L$ is equal to  $Q$ for all $L \ge L_0$, for large $L$, the multiplicity of $\tau^L_q$ is the same as $\tau_q$.  Proposition 2.3 (iii) of \cite{chatelin1981} can then be applied to conclude that there exists a set of eigenfunctions $\theta^L_q$ of $T_L$ such that $\theta^L_q \rightarrow \theta_q$.

\bibliographystyle{chicago}
\bibliography{LCDAbib}

\end{document}